\documentclass[
 amsmath,amssymb,prd
]{revtex4-2}

\usepackage{graphicx}
\usepackage{dcolumn}
\usepackage{bm}

\usepackage{mathtools,amssymb,amsthm}
\usepackage{enumerate}
\usepackage{enumitem}
\usepackage{bbm}

\usepackage{tensor}

\usepackage{graphicx}
\usepackage{upgreek}

\usepackage[main=british,ngerman]{babel}
\usepackage[T1]{fontenc}
\usepackage[utf8]{inputenc}  
\usepackage[utf8]{luainputenc}
\usepackage{comment}

\usepackage{datetime}
\newdateformat{monthyeardate}{%
  \monthname[\THEMONTH] \THEYEAR}

\usepackage{caption}
\usepackage{subcaption}

\usepackage[autostyle=true,english=american]{csquotes}

\usepackage[dvipsnames]{xcolor}
\definecolor{darkgreen}{rgb}{0,0.7,0}
\usepackage[ocgcolorlinks=true]{hyperref}
\hypersetup{
	final,
	unicode=true,
	linkcolor={red!50!black},
	citecolor={green!50!black},
	urlcolor={blue!50!black},
	pdftitle={GHY term in (S)TEGR},
	pdfauthor={Bastian He\ss{}},
	pdfdisplaydoctitle=true,
}
\usepackage{bookmark}

\usepackage{cancel}

\renewcommand{\d}{\mathrm{d}}

\newcommand{\onehalf}{\frac{1}{2}}

\renewcommand{\L}{\mathcal{L}}
\newcommand{\M}{\mathcal{M}}

\newcommand*{\defeq}{\mathrel{\vcenter{\baselineskip0.5ex \lineskiplimit0pt
			\hbox{\scriptsize.}\hbox{\scriptsize.}}}%
	=}
	
\newcommand{\hodge}{*}
\newcommand{\interior}{\rfloor}

\renewcommand{\overcirc}[1]{\mathring{#1}}

\newcommand{\n}{\mathbf{n}}

\newcommand{\mcomma}{\, ,}
\newcommand{\mdot}{\, .}

\begin{document}

\preprint{APS/123-QED}

\title{Gibbons-Hawking-York boundary  terms and\\ the generalized geometrical trinity of gravity}

\author{Johanna Erdmenger}
\author{Bastian Heß}%
 \email{bastian.hess@uni-wuerzburg.de}
\affiliation{%
 Institute for Theoretical Physics and Astrophysics, Julius-Maximilians-Universität Würzburg, Am~Hubland, D-97074~Würzburg, Germany
}%
\affiliation{
 Würzburg-Dresden Cluster of Excellence on Complexity and Topology in Quantum Matter ct.qmat
}

\author{Ioannis Matthaiakakis}
\affiliation{
 Dipartimento di Fisica, Universit\`{a} di Genova, via Dodecaneso 33, I-16146 Genova, Italy
}%
\affiliation{
 I.N.F.N. - Sezione di Genova, via Dodecaneso 33, I-16146 Genova, Italy
}
\affiliation{
Mathematical Sciences and STAG Research Centre, University of Southampton, Highfield,
Southampton SO17 1BJ, United Kingdom}%
\author{René Meyer}
\affiliation{%
 Institute for Theoretical Physics and Astrophysics, Julius-Maximilians-Universität Würzburg, Am~Hubland, D-97074~Würzburg, Germany
}%
\affiliation{
 Würzburg-Dresden Cluster of Excellence on Complexity and Topology in Quantum Matter ct.qmat
}

\date{Monday 2$^\mathrm{nd}$ September, 2024}

\begin{abstract}
General relativity (GR) as described in terms of curvature by the Einstein-Hilbert action is dynamically equivalent to theories of gravity formulated in terms of spacetime torsion or non-metricity. This forms what is called the geometrical trinity of gravity. The theories corresponding to this trinity are, apart from GR, the teleparallel (TEGR) and symmetric teleparallel (STEGR) equivalent theories of general relativity, respectively, and their actions are equivalent to GR up to boundary terms $B$. We investigate how the Gibbons-Hawking-York (GHY) boundary term of GR changes under the transition to TEGR and STEGR within the framework of metric-affine gravity. We show that $B$ is the difference between the GHY term of GR and that of metric-affine gravity. Moreover, we show that the GHY term for both TEGR and STEGR must vanish for consistency of the variational problem. Furthermore, our approach allows to extend the equivalence between GR, TEGR and STEGR beyond the Einstein-Hilbert action to any action built out of the curvature two-form, thus establishing the generalized geometrical trinity of gravity. We argue that these results will be particularly useful in view of studying  gauge/gravity duality for theories with torsion and non-metricity. 
\end{abstract}

\maketitle

\noindent\rule{\textwidth}{1pt}
\pdfbookmark[section]{\contentsname}{toc}
\vspace{-10pt}
\tableofcontents
\vspace{10pt}
\noindent\rule{\textwidth}{1pt}
	
	\section{Introduction}
\label{sec_introduction}

Einstein's theory of general relativity (GR) is a theory with tremendous success in describing gravity~\cite{Weinberg:1972kfs,hawking1973,wald2010}. In standard formulations of GR, gravity is described entirely in terms of spacetime curvature. However, torsion as well as non-metricity are two additional geometric quantities any given spacetime may exhibit~\cite{hehl1995}. In recent years, progress has been made towards understanding theories of gravity including torsion and non-metricity, as well as their implications to cosmology and black hole physics (see e.g.~\cite{hess2022,bahamonde2021_book,bahamonde2021,paci2021,saridakis2021,aldrovandi2013,boehmer2021}). Our motivation is to examine theories of gravity containing torsion and non-metricity within the context of gauge/gravity duality. Within gauge/gravity duality, a theory of gravity in an (asymptotically local) Anti-de Sitter spacetime provides us with a dual description of the dynamics of a strongly coupled conformal field theory (CFT)~\cite{maldacena1999,witten1998,gubser1998,erdmenger2015}. In addition this dual CFT is realized holographically, in the sense that it is defined on the boundary of the spacetime in which gravity propagates. Gauge/gravity duality has been used successfully to analyze extensively the transport properties of strongly coupled CFTs (see e.g.~\cite{PhysRevLett.94.111601,PhysRevD.75.085020,Charmousis2010, PhysRevB.98.195143,disante2019}). However, these analyses have been performed almost entirely in torsion-free, metric curved spacetimes~\footnote{See~\cite{leigh2008,Ciambelli:2019bzz,iosifidis2018,guersoy2020} for some early work on torsionful and non-metric realizations of gauge/gravity duality.}. We aim at introducing torsion and non-metricity into this duality in order to describe spin and hypermomentum transport in strongly coupled systems. In particular, we want to understand the hydrodynamic transport of spin and hypermomentum in strongly coupled fluids, such as Dirac and Weyl semimetals or the quark gluon plasma~\cite{erdmenger2008,Ammon:2017ded,bardarson2019,Ammon:2020rvg,Cartwright:2021qpp,Ammon:2021pyz,Hongo:2022izs,xian2022}, by using the fluid/gravity correspondence~\cite{hubeny2011,Bhattacharyya:2007vjd,rangamani2009}.
Since the spin and hypermomentum of the dual fluid are encoded on the boundary of the spacetime by definition, it is important to study the gravitational boundary terms involved in torsionful and non-metric theories of gravity.

Theories in which either torsion or non-metricity are non-vanishing while curvature vanishes identically are called teleparallel and symmetric teleparallel theories of gravity, respectively. We focus on a subset of these theories, the teleparallel equivalent of general relativity (TEGR)~\cite{bahamonde2021_book,bahamonde2021,paci2021,saridakis2021,aldrovandi2013,Heisenberg:2022mbo,Hohmann:2022mlc} and the symmetric teleparallel equivalent of general relativity (STEGR)~\cite{nester1998,Adak:2004uh,Adak:2005cd,Adak:2008gd,BeltranJimenez:2017tkd,jarv2018,Hohmann:2021rmp}. As the names suggest, these theories are dynamically equivalent to GR. A particular advantage of these reformulations of GR is the fact that their dynamics are described by Lagrangians reminiscent of gauge theory Lagrangians. For example, model Lagrangians for (S)TEGR are
\begin{align}\label{eq_gauge_lagrangians}
    \L_\mathrm{TEGR}\propto T_\mu\wedge\star T^\mu\mcomma
    \qquad
    \L_\mathrm{STEGR}\propto Q^{\mu\nu}\wedge\star Q_{\mu\nu}\mcomma
\end{align}
where $T^\mu$ and $Q_{\mu\nu}$ are the differential forms of torsion and non-metricity, respectively~\footnote{Beyond~\eqref{eq_gauge_lagrangians}, we may study actions built out of the weighted sums of the irreducible decompositions of $T^\mu$ and $Q_{\mu\nu}$ under $\mathrm{SO}(1,n-1)$, see~\cite{hehl1995,adak2006}.}.

\noindent There are several points in relation to actions of the form~\eqref{eq_gauge_lagrangians} that we would like to emphasize. First, the gauge degrees of freedom of $T^\mu$ and $Q_{\mu\nu}$ are defined on vector spaces that are locally isomorphic to the underlying spacetime. As a result, the standard Hodge dual used in defining gauge theories with internal degrees of freedom must be modified in non-trivial ways in order to take into account their action on the gauge indices~\cite{lucas2008,aldrovandi2013}. In particular, it is necessary to define a generalized Hodge dual whose explicit expression is not known for all dimensions~\cite{huang2014}. As a result, it is not obvious how to write down the most general (S)TEGR Lagrangians in dimensions different than four. Hence, the gauge theory-like Lagrangians~\eqref{eq_gauge_lagrangians} have only limited use in arbitrary dimensions, and for explicit calculations. A version of the (S)TEGR Lagrangian without generalized Hodge duals is needed~\footnote{Note, however, that a way to circumvent this issue is implicitly present in~\cite{nester1998,BELTRANJIMENEZ2020135422}, where the authors discuss (S)TEGR in $3+1$ dimensions. Our analysis generalizes their approach to a broader class of Lagrangians.}.

The second feature of (S)TEGR we focus on is its boundary term. It is well-known that GR has a well-defined variational formulation in spacetimes with boundaries only if a boundary Gibbons-Hawking-York (GHY) term is included alongside the bulk Einstein-Hilbert Lagrangian~\cite{york1972,gibbons1977,wald2010,hess2022}. A so far unsolved issue is whether a GHY equivalent term must be included alongside the bulk actions in~\eqref{eq_gauge_lagrangians} (see e.g.~\cite{PhysRevD.96.044042,Heisenberg:2018vsk,jimenez2019,paci2021,Capozziello:2022zzh}). 

Finally, we note that the geometrical trinity of gravity was established for the Einstein-Hilbert action~\cite{Heisenberg:2018vsk,jimenez2019,Capozziello:2022zzh}~\footnote{See also \cite{Nakayama:2022qbs,Wolf:2023rad,March:2023trv} for extensions of the geometrical trinity to unimodular and non-relativistic gravity, as well as \cite{Iosifidis:2023eom} for a discussion of the trinity in the presence of matter.}. However, extensions of GR, including arbitrary polynomials of the Riemann tensor, are routinely studied in the literature, see e.g.~\cite{erdmenger2015}. It is then necessary to understand whether such extensions may be equivalently recast as extensions of (S)TEGR. 

In the present paper, we use the formalism developed in~\cite{hess2022} to address all three of these points. In particular, we consider GR from the point of view of metric-affine gravity (MAG)~\cite{hehl1995}. MAG describes the contributions of both torsion and non-metricity to curvature in terms of a single geometric object, the deformation one-form $A\indices{^\mu_\nu}$ defined in~\eqref{eq_c_split_t_q}. This allows us to re-derive TEGR and STEGR simultaneously. We show that the boundary term generated when transitioning from GR to one of its equivalents, simply translates the GHY term of GR to the GHY term of MAG generated by the curvature two-form. Since curvature is set to zero when we enforce teleparallelism, we conclude that (S)TEGR does \textit{not} need a boundary term. In addition, our formalism is defined naturally in any dimension. Thus, our derivation of the geometrical trinity also provides the Lagrangian of (S)TEGR in general dimensions, see~\eqref{eq_final_action}. 

Finally, we investigate under which conditions we may generalize the equivalence between GR and (S)TEGR to extended theories of gravity. We show that the geometrical trinity of gravity exists for \textit{any} GR extension that may be written in terms of the curvature two-form alone. In contrast to the ordinary geometrical trinity, the boundary term for our generalized trinity does not vanish identically. Instead it is given by a specific expression, which for topological gravity takes the form of a Chern-Simons gauge theory for the deformation one-form, defined as the difference between the full and the Christoffel connections. This boundary term in not necessary for the variational problem to be well-defined. Instead, in general, it imposes boundary conditions on the bulk torsion or non-metricity in order for the action to be equivalent to the corresponding GR extension.

In section~\ref{sec_setup}, we begin our investigation by setting up the geometric notions necessary to understand our results. In section~\ref{sec_main}, we then derive the equivalence between GR and (S)TEGR, as well as their corresponding GHY terms, from the point of view of MAG via a differential geometric formalism. A generalization of our proof to higher-curvature extensions of GR is given in section~\ref{sec_generalization}. Finally, in section~\ref{sec_conclusions} we summarize our results and provide some outlook for further research. We express our results in a more familiar tensor calculus notation in appendix~\ref{sec_tensor}.

\section{Geometric setup}
\label{sec_setup}
In this section we define the basic geometric objects necessary for understanding the derivation of our main result in sections~\ref{sec_main} and~\ref{sec_generalization}. For a more thorough discussion of the formalism employed, the reader may consult~\cite{hess2022}. 

We begin by considering an $n$-dimensional manifold~$\M$, the spacetime where our theory of gravity lives on. We assume $\M$ is equipped with a connection one-form~$\omega^\mu_\nu=\Gamma^\mu_{\rho\nu}\theta^\rho$ and a metric $\d s^2=g_{\mu\nu}\theta^\mu\otimes\theta^\nu$, as well as a coframe basis $\theta^\mu$ of the cotangent spaces of $\M$, which is not necessarily holonomic ($\d\theta^\mu \neq 0$)~\footnote{Greek indices take values in $\{0,\dots,n-1\}$.}. The fields $\omega^\mu_\nu$, $g_{\mu\nu}$ and $\theta^\mu$ are the dynamical fields of the theories of gravity we consider. The gravitational Lagrangians we focus on are those built out of the field strengths of the aforementioned dynamical fields defined as
\begin{equation}\label{eq_def_field_strengths}
        \begin{aligned}
            &\text{the curvature two-form} \vphantom{\onehalf}
            \\
            &\text{the torsion two-form} \vphantom{\onehalf}
            \\
            &\text{the non-metricity one-form} \vphantom{\onehalf}
        \end{aligned}\quad
        \begin{aligned}
            &\Omega\indices{^\mu_\nu}\defeq \d \omega^\mu_\nu+\omega^\mu_\rho\wedge\omega^\rho_\nu =\onehalf R\indices{^\mu_\nu_\rho_\sigma}\theta^\rho\wedge\theta^\sigma\mcomma 
            \\
            &T^\mu\defeq D \theta^\mu = \d\theta^\mu + \omega^\mu_\nu\wedge\theta^\nu =\onehalf T\indices{^\mu_\rho_\sigma}\theta^\rho\wedge\theta^\sigma\mcomma
            \\
            &Q_{\mu\nu}\defeq -D g_{\mu\nu}
            = -(\d g_{\mu\nu} -\omega^\rho_\mu g_{\rho\nu}-\omega^\rho_\nu g_{\mu\rho})
            =Q_{\mu\nu\rho}\theta^\rho
            \mcomma \vphantom{\onehalf}
        \end{aligned}
\end{equation}
where $\d$ denotes the exterior derivative and $D$ the exterior covariant derivative with respect to the connection~$\omega^\mu_\nu$~\footnote{Note that we chose to write the tensor components of~\eqref{eq_def_field_strengths} in terms of $\theta^\mu$ instead of a co-ordinate basis $dx^I$ in order to work with only one index type for simplicity. This is always possible since the components of $\theta^\mu = \theta^\mu_I dx^I$ provide a soldering/intertwiner form connecting the two bases.}. The components of the curvature two-form are those of the Riemann tensor~$R\indices{^\mu_\nu_\rho_\sigma}$, while those of torsion and non-metricity are the torsion tensor~$T\indices{^\mu_\rho_\sigma}$ and non-metricity tensor~$Q_{\mu\nu\rho}$, respectively.

We define GR, TEGR and STEGR in terms of $\Omega\indices{^\mu_\nu}$, $T^\mu$ and $Q_{\mu\nu}$ as 
\begin{equation}\label{eq_def_field_theories}
        \begin{aligned}
            &\text{GR} \vphantom{\onehalf}
            \\
            &\text{TEGR} \vphantom{\onehalf}
            \\
            &\text{STEGR} \vphantom{\onehalf}
        \end{aligned}\quad
        \begin{aligned}
            &\Omega\indices{^\mu_\nu}\neq 0 \mcomma~T^\mu = 0\mcomma~Q_{\mu\nu}= 0\mcomma \vphantom{\onehalf} 
            \\
            &\Omega\indices{^\mu_\nu}= 0 \mcomma~T^\mu\neq 0\mcomma~Q_{\mu\nu}= 0\mcomma \vphantom{\onehalf}
            \\
            &\Omega\indices{^\mu_\nu}= 0\mcomma~T^\mu= 0\mcomma~Q_{\mu\nu}\neq 0
            \mdot \vphantom{\onehalf}
        \end{aligned}
\end{equation}
The definition~\eqref{eq_def_field_theories} also clarifies the naming of (S)TEGR as teleparallel theories: Both TEGR and STEGR have by definition vanishing curvature, meaning a vector parallel transported by $D$ in those theories remains parallel to itself. In addition, STEGR is symmetric in the sense that the fundamental dynamical field, $Q_{\mu\nu}$, is symmetric in its two indices unlike the two-form $T^\mu$ in TEGR. 

GR is special within this family of theories, because the vanishing of both~$T^\mu$ and~$Q_{\mu\nu}$ constrains the connection~$\omega^\mu_\nu$ to be the Christoffel connection: If we denote the Christoffel connection by $\overcirc{\omega}^\mu_\nu$, its corresponding covariant derivative, $\overcirc{D}$, is the unique covariant derivative satisfying
\begin{subequations}\label{eq_Christoffel_conditions}\begin{align}
    0&=\overcirc{T}^\mu=\overcirc{D}\theta^\mu=\d\theta^\mu+\overcirc{\omega}^\mu_\nu\wedge\theta^\nu \quad\Leftrightarrow\quad \d\theta^\mu=-\overcirc{\omega}^\mu_\nu\wedge\theta^\nu\mcomma
    \\
    0&= \overcirc{Q}_{\mu\nu}=-\overcirc{D}g_{\mu\nu}=-\d g_{\mu\nu}+\overcirc{\omega}^\rho_\mu g_{\rho\nu}+\overcirc{\omega}^\rho_\nu g_{\mu\rho}\mdot
\end{align}\end{subequations}
In discussing (S)TEGR it is useful to isolate the part of the connection that allows for non-trivial torsion and non-metricity. We do this by defining the deformation one-form
\begin{align}\label{eq_def_c}
    A\indices{^\mu_\nu}\defeq \omega^\mu_\nu-\overcirc{\omega}^\mu_\nu\mcomma
\end{align}
with respect to which we may express $T^\mu$ and $Q_{\mu\nu}$ as
\begin{align}\label{eq_c_torsion_non-metricity}
    A_{\mu\nu}+A_{\nu\mu}&=Q_{\mu\nu}\mcomma
    \qquad
    A\indices{^\mu_\nu}\wedge \theta^\nu= T^\mu\mdot
\end{align}
The latter equations do not imply that the antisymmetric part of $A_{\mu\nu}$ consists of torsion only. In fact, we have  $(A_{\mu\nu}-A_{\nu\mu})\wedge\theta^\nu=2T_\mu-Q_{\mu\nu}\wedge\theta^\nu$. This will be crucial when discussing the STEGR boundary term, as this anti-symmetric part is how non-metricity enters the equations. 

To give more intuition for the properties of $A_{\mu\nu}$, we invert~\eqref{eq_c_torsion_non-metricity} and express $A_{\mu\nu}$ in terms of torsion and non-metricity~\cite{hehl1995} as
\begin{align}\label{eq_c_split_t_q}
    A_{\mu\nu}=-\vartheta_{[\mu}\interior T_{\nu]}+\onehalf\left(\vartheta_\mu\interior \vartheta_\nu\interior T_\rho\right)\theta^\rho+\onehalf Q_{\mu\nu}+\left(\vartheta_{[\mu}\interior Q_{\nu]\rho}\right)\theta^\rho\mcomma
\end{align}
where $\interior$ is the interior product and $\vartheta_\mu$ is the dual frame to $\theta^\mu$, $\vartheta_\nu\interior\theta^\mu = \delta^\mu_\nu$. In terms of components, we have 
\begin{align}
\label{eq_A_components}
    A_{\mu\nu\rho}
    =g_{\mu\sigma}\left( \Gamma^\sigma_{\rho\nu} - \overcirc{\Gamma}^\sigma_{\rho\nu}\right)
    =\onehalf \left( -T_{\rho\nu\mu}+T_{\nu\mu\rho}+T_{\mu\rho\nu} \right)
    +\onehalf\left( -Q_{\rho\nu\mu}+Q_{\nu\mu\rho}+Q_{\mu\rho\nu} \right)\mdot
\end{align}
We observe that the torsion terms define the contorsion tensor $K_{\mu\nu\rho}$, well-known in Einstein-Cartan theories of gravity~\cite{cartan1922,cartan1923,cartan1924,cartan1925,sciama1964,kibble1960,trautman2006}, while the non-metricity terms define a Weyl connection in the special case where $Q_{\mu\nu} = Q g_{\mu\nu}$ for some one-form $Q$ (see e.g.~\cite{Ciambelli:2019bzz}). Furthermore, from~\eqref{eq_c_split_t_q} we see that $A_{\mu\nu}$ captures any deviation from GR, be it due to torsion or non-metricity. Therefore, the deformation one-form allows us to describe both TEGR and STEGR simultaneously within a single formalism.

To complete our discussion of geometric preliminaries, we introduce a boundary $\partial \M $ in our spacetime $\M$, as well as the decomposition of the dynamical fields in contributions normal and tangent to $\partial \M$. Again, a thorough discussion of what follows along with proofs may be found in~\cite{hess2022} and the references therein.
The geometry of $\partial \M$ is described by tangent vectors $e^a_\mu$ and a normal vector~$n^\mu$, fulfilling $e^a_\mu n^\mu=0$. Latin indices denote internal indices of the boundary manifold and take values in $\{0,\dots,n-2\}$. We normalize $n^\mu$ as $n_\mu n^\mu=\varepsilon$, where $\varepsilon=-1$ for spacelike and $\varepsilon=+1$ for timelike boundaries. The connection one-form on the boundary is induced by the connection in the bulk via
\begin{align}
    \label{eq_connection_transformation}
    \omega^b_a=e^b_\mu\left(\d e^\mu_a+\omega^\mu_\nu e^\nu_a\right)\mcomma
\end{align}
whereas general tensors are pulled back to $\partial\M$ as usual by contraction with $e^a_\mu$. In this manner, $\phi^a\defeq e^a_\mu\theta^\mu$ defines a basis coframe on $\partial\M$ and $\gamma_{ab}\defeq e^\mu_{a\vphantom{b}} e^{\nu\vphantom{\mu}}_b g_{\mu\nu}$ a boundary metric. These pullbacks allow us to decompose $\theta^\mu$ and $g_{\mu\nu}$ into tangent and normal contributions as
\begin{align}\label{eq_g_decomposition}
    \theta^\mu&=e^\mu_a\phi^a+N n^\mu\phi\mcomma
    \qquad
    g_{\mu\nu}=e^a_\mu e^b_\nu \gamma_{ab}+\varepsilon n_\mu n_\nu\mcomma
\end{align}
where $N\defeq1/\sqrt{\left|n^\mu n_\mu\right|}$ is the normalization factor of $n^\mu$ and $\phi\defeq\frac{\varepsilon}{N}n_\mu\theta^\mu$. 
The embedding of~$\partial\M$ in $\M$ is characterized by the extrinsic curvature one-forms~\footnote{We denote $K^a$ and $\tilde{K}_a$ as extrinsic curvatures, since they reduce to the usual definition in the Riemannian case. Their geometric interpretation, however, might be more involved. We thank Martin Krššák for pointing this out.}
\begin{align}\label{eq_extrinsic_curvature}
    K^a\defeq e^a_\mu D n^\mu\qquad\text{and}\qquad\tilde{K}_a\defeq e^\mu_a D n_\mu = K_a - e^\mu_a n^\nu Q_{\mu\nu} \mcomma
\end{align}
which differ only due to non-metricity. The projections of the components of these one-forms to the boundary give the well-known expressions of extrinsic curvature in tensor language~\cite{hess2022}.

\section{(S)TEGR action and its boundary term in differential geometry}\label{sec_main}

In the present section, we present our derivation of the GHY boundary term for (S)TEGR. The derivation boils down to two steps, first we consider the Einstein-Hilbert Lagrangian of GR and use~\eqref{eq_def_c} to express it in terms of $A\indices{^\mu_\nu}$ and $\Omega\indices{^\mu_\nu}$. Second, we enforce teleparallelism by setting $\Omega\indices{^\mu_\nu} = 0$. During the first part of this derivation a boundary term appears which we must treat carefully as it is this boundary term that holds all the information regarding the GHY term for (S)TEGR.

To begin our derivation, we insert the deformation one-form~\eqref{eq_def_c} into the curvature definition~\eqref{eq_def_field_strengths} to obtain
\begin{align}\label{eq_curvature_split}
    \Omega\indices{^\mu_\nu}=\overcirc{\Omega}\indices{^\mu_\nu}+\overcirc{D}A\indices{^\mu_\nu}+A\indices{^\mu_\rho}\wedge A\indices{^\rho_\nu}\mcomma
\end{align}
where $\overcirc{\Omega}\indices{^\mu_\nu}\defeq \d \overcirc{\omega}^\mu_\nu+\overcirc{\omega}^\mu_\rho\wedge\overcirc{\omega}^\rho_\nu$ is the curvature two-form built from the Christoffel connection~$\overcirc{\omega}^\mu_\nu$, that is the curvature two-form of GR. For the Einstein-Hilbert action in generic $n$~dimensions, this implies
\begin{subequations}\label{eq_eh_action}
\begin{align}
     S^\mathrm{EH}_{\overcirc{\Omega}}&=\frac{1}{2\kappa}\int\d^nx\sqrt{-g}\overcirc{R}+ S^\mathrm{EH}_{\overcirc{\Omega},\mathrm{GHY}}=\frac{1}{2\kappa}\int\eta^{\mu\nu}\wedge\overcirc{\Omega}_{\mu\nu}+ S^\mathrm{EH}_{\overcirc{\Omega},\mathrm{GHY}}
    \\\label{eq_eh_action_2nd_line}&=
    \frac{1}{2\kappa}\int\eta^{\mu\nu}\wedge\left(\Omega_{\mu\nu}-\overcirc{D}A\indices{_\mu_\nu}-A\indices{_\mu_\rho}\wedge A\indices{^\rho_\nu}\right)+ S^\mathrm{EH}_{\overcirc{\Omega},\mathrm{GHY}}
    \mcomma
\end{align}
\end{subequations}
where we define~$\eta^{\mu\nu}\defeq\hodge(\theta^\mu\wedge\theta^\nu)$ in which $\hodge$ denotes the Hodge duality. We supplement the action by the well-known GHY term of GR, given by
\begin{align}\label{eq_Riemannian_GHY}
    S^\mathrm{EH}_{\overcirc{\Omega},\mathrm{GHY}}=
    \frac{\varepsilon}{\kappa}\int_{\partial\M}\left.\overcirc{K}_a\wedge\eta^{\n a}\right|_{\partial\mathcal{M}}
    =\frac{\varepsilon}{\kappa}\int_{\partial\mathcal{M}}\d \text{Vol}_{\partial\mathcal{M}} \sqrt{|\gamma|}\overcirc{K}\indices{^a_a}\mcomma
\end{align}
where the extrinsic curvature one-form was defined in~\eqref{eq_extrinsic_curvature} (for $Q_{\mu\nu}=0$) and we abbreviate $\eta^{\n a}\defeq n_\mu e^a_\nu \eta^{\mu\nu}$.  Recall that the GHY term is necessary to obtain a well-defined variational problem~\cite{york1972,gibbons1977,wald2010}. According to~\eqref{eq_def_field_theories}, the action in~\eqref{eq_eh_action_2nd_line} represents TEGR if $Q_{\mu\nu}=0=\Omega_{\mu\nu}$, in which case $A\indices{^\mu_\nu}$ is called the contorsion one-form, while it represents STEGR if we choose $T_\mu=0$ and $\Omega_{\mu\nu}=0$ instead.

Next, we consider the $A_{\mu\nu}$ terms in~\eqref{eq_eh_action}. In particular, we focus on $\overcirc{D}A_{\mu\nu}$, which is the all important boundary term mentioned earlier. To see this explicitly, we isolate this term and consider the action 
\begin{align}\label{eq_Dc_action_1st_def}
    S^\mathrm{\overcirc{D}A}\defeq -\frac{1}{2\kappa}\int_\M\eta^{\mu\nu}\wedge\overcirc{D}A\indices{_\mu_\nu}
    =-\frac{1}{2\kappa}\int_\M A_{\mu\nu}\wedge\overcirc{D}\eta^{\mu\nu}-\frac{1}{2\kappa}\int_{\partial\M}\left.A_{\mu\nu}\wedge\eta^{\mu\nu}\vphantom{\left(\tilde{K}_a\right)}\right|_{\partial\M}
    \mdot
\end{align}
The first term on the right-hand side of~\eqref{eq_Dc_action_1st_def} vanishes identically after we recall the definitions of the Hodge star operator~$\hodge$ and the Christoffel exterior covariant derivative~$\overcirc{D}$ from~\eqref{eq_Christoffel_conditions}. These definitions imply
\begin{align}
    \overcirc{D}\eta^{\mu\nu}=\overcirc{D}\left(\frac{\sqrt{|\det g|}}{(n-2)!}\varepsilon_{\sigma_1\dots\sigma_2\rho_1\dots\rho_{n-2}}g^{\sigma_1\mu}g^{\sigma_2\nu}\theta^{\rho_1}\wedge\dots\wedge\theta^{\rho_{n-2}}\right) =0\mcomma
\end{align}
such that only the boundary term remains in $S^\mathrm{\overcirc{D}A}$,
\begin{align}\label{eq_Dc_action}
    S^\mathrm{\overcirc{D}A}
    =-\frac{1}{2\kappa}\int_{\partial\M}\left.A_{\mu\nu}\wedge\eta^{\mu\nu}\vphantom{\left(\tilde{K}_a\right)}\right|_{\partial\M}
    \mdot
\end{align}
To proceed, we need to express $S^\mathrm{\overcirc{D}A}$ entirely in terms of boundary data. To achieve this, we use the definition of $A\indices{^\mu_\nu}= \omega^{\mu}_\nu - \overcirc{\omega}^\mu_\nu$ as well as the $3+1$ decomposition of the connection~\footnote{We follow convention and call the decomposition of the connection in tangential and normal components the connection's $3+1$ decomposition. However, our results are valid in any dimension.}. We derive said decomposition by employing
\begin{align}\label{eq_delta_split}
    \delta^\mu_\nu=e^\mu_a e^a_\nu+\varepsilon n^\mu n_\nu
\end{align}
to decompose the indices of $\omega^\mu_\nu$ into boundary normal and tangent contributions. Namely, 
\begin{align}\label{eq_connection_decomposition}
    \omega^\mu_\nu
    =\delta^\mu_\alpha \delta^\beta_\nu\omega^\alpha_\beta 
    =e^\mu_a e^b_\nu (e^a_\alpha e_b^\beta \omega^\alpha_\beta)
    +n^\mu n_\nu (n_\alpha n^\beta \omega^\alpha_\beta)
    +\varepsilon n^\mu e^a_\nu (n_\alpha e_a^\beta \omega^\alpha_\beta)
    +\varepsilon e^\mu_a n_\nu (e_\alpha^a n^\beta \omega^\alpha_\beta)\mdot
\end{align}
The decomposition of $\omega^\mu_\nu$ then amounts to determining the projections in parentheses in~\eqref{eq_connection_decomposition}. 
The projection of $\omega^\mu_\nu$ on $\partial M$ in both indices is given in~\eqref{eq_connection_transformation} as
\begin{align}\label{eq_decomposition_ee}
    e^a_\mu e^\nu_b \omega^\mu_\nu= \omega^a_b - e^a_\mu\d e^\mu_b\mdot
\end{align}
The corresponding projections in which one of the indices is projected to the normal vector~$n_\mu$ instead are obtained from the extrinsic curvatures $K^a$ and $\tilde{K}_a$ in~\eqref{eq_extrinsic_curvature}, which decompose as
\begin{align}
    K^a\defeq e^a_\mu D n^\mu=e^a_\mu \d n^\mu+e^a_\mu \omega^\mu_\nu n^\nu
    \qquad\text{and}\qquad
    \tilde{K}_a\defeq e^\mu_a D n_\mu = e^\mu_a \d n_\mu-e^\mu_a \omega^\nu_\mu n_\nu\mdot
\end{align}
We solve for the projections of the connection as
\begin{align}\label{eq_decomposition_ne}
    e^a_\mu n^\nu\omega^\mu_\nu=K^a-e^a_\mu\d n^\mu
    \qquad\text{and}\qquad
    n_\mu e^\nu_a \omega^\mu_\nu = -\tilde{K}_a+e^\mu_a\d n_\mu\mdot
\end{align}
Lastly, the twice normal projected part of the connection is obtained from the normal component of non-metricity,
\begin{align}
    Q_{\n\n}\defeq n^\mu n^\nu Q_{\mu\nu} =2n_\mu D n^\mu =2n_\mu \d n^\mu+2n_\mu \omega^\mu_\nu n^\nu \mcomma
\end{align}
yielding
\begin{align}\label{eq_decomposition_nn}
    n_\mu n^\nu \omega^\mu_\nu = \onehalf Q_{\n\n}-n_\mu\d n^\mu\mdot
\end{align}
We thus obtain the $3+1$ decomposition of the connection as 
\begin{align}\begin{split}
\label{eq_connection_split}
    \omega^\mu_\nu=&e^\mu_a e^b_\nu(\omega^a_b-e^a_\rho\d e^\rho_b)+n^\mu n_\nu(\onehalf Q_{\n\n}-n_\rho\d n^\rho)
    \\&+\varepsilon n^\mu e^a_\nu(e^\rho_a\d n_\rho-\tilde{K}_a)+\varepsilon e^\mu_a n_\nu(K^a-e^a_\rho\d n^\rho)
\end{split}\end{align}
which immediately implies the 3+1 decomposition of the deformation one-form~\eqref{eq_def_c}~\footnote{The $3+1$ decomposition of $\overcirc{\omega}^\mu_\nu$ is given by~\eqref{eq_connection_split} after setting $Q_{\n\n} = 0= \tilde{K}^a - K^a$.}
\begin{align}\label{eq_c_decomposition}
    A\indices{^\mu_\nu}
    =e^\mu_a e^b_\nu A\indices{^a_b}+\onehalf n^\mu n_\nu Q_{\n\n}+\varepsilon n^\mu e^a_\nu(\overcirc{K}_a-\tilde{K}_a)+\varepsilon e^\mu_a n_\nu (K^a-\overcirc{K}^a)\mcomma
\end{align}
where $A\indices{^a_b}\defeq \omega^a_b-\overcirc{\omega}^a_b$ is the boundary deformation one-form. Knowing this decomposition, we come back to the boundary action~\eqref{eq_Dc_action} which evaluates to
\begin{align}\label{eq_boundary_action_intermediate}
    S^\mathrm{\overcirc{D}A}=
    -\frac{1}{2\kappa}
    \int_{\partial\M}\left. A\indices{_a_b}\wedge\left(e^a_\mu e^b_\nu\eta^{\mu\nu}\right)\right|_{\partial\mathcal{M}}
    -\frac{\varepsilon}{2\kappa}\int_{\partial\M}\left.\left(2\overcirc{K}_a-\tilde{K}_a-K_a\right)\wedge\eta^{\n a}\right|_{\partial\mathcal{M}}\mdot
\end{align}
We further simplify this boundary term by examining the Hodge dual in $\left. A\indices{_a_b}\wedge\left(e^a_\mu e^b_\nu\eta^{\mu\nu}\right)\right|_{\partial\mathcal{M}}$. Using the definition $\eta^{\mu\nu}=\hodge\left(\theta^\mu\wedge\theta^\nu\right)$ we obtain
\begin{align}\label{eq_pullback_ab_term}
    \left. A\indices{_a_b}\wedge\left(e^a_\mu e^b_\nu\eta^{\mu\nu}\right)\right|_{\partial\mathcal{M}}
    =
    \frac{\sqrt{|\det g|}}{(n-2)!}\varepsilon_{\mu\nu\rho_1\dots\rho_{n-2}}e^\mu_a e^\nu_b e^{\rho_1}_{c_1}\cdots e^{\rho_{n-2}}_{c_{n-2}}A^{ab}\wedge\phi^{c_1}\wedge\dots\phi^{c_{n-2}}\mcomma
\end{align}
where $\phi^a=e^a_\mu\theta^\mu$ is the boundary coframe. Note that all $n$ indices of the $\varepsilon$-symbol take their values on the $(n-1)$-dimensional boundary. As a result, $\left. A\indices{_a_b}\wedge\left(e^a_\mu e^b_\nu\eta^{\mu\nu}\right)\right|_{\partial\mathcal{M}}$ vanishes identically. This simplifies the boundary action~\eqref{eq_boundary_action_intermediate} to
\begin{align}
\label{eq_DA_EH}
    S^\mathrm{\overcirc{D}A}= 
    -\frac{\varepsilon}{2\kappa}\int_{\partial\M}\left.\left(2\overcirc{K}_a-\tilde{K}_a-K_a\right)\wedge\eta^{\n a}\right|_{\partial\mathcal{M}}\mdot
\end{align}
Inserting everything into the Einstein-Hilbert action~\eqref{eq_eh_action}, we obtain
\begin{align}\label{eq_eh_action_transformed}\begin{split}
     S^\mathrm{EH}_{\overcirc{\Omega}}
    &=\frac{1}{2\kappa}\int_\M\eta^{\mu\nu}\wedge\overcirc{\Omega}_{\mu\nu}+\frac{\varepsilon}{\kappa}\int_{\partial\M}\left.\overcirc{K}_a\wedge\eta^{\n a}\right|_{\partial\mathcal{M}}
    \\
    &=
    \frac{1}{2\kappa}\int_\M\eta^{\mu\nu}\wedge\left(\Omega_{\mu\nu}-A\indices{_\mu_\rho}\wedge A\indices{^\rho_\nu}\right)+\frac{\varepsilon}{\kappa}\int_{\partial\M}\left.\overcirc{K}_a\wedge\eta^{\n a}\right|_{\partial\mathcal{M}}
    +S^\mathrm{\overcirc{D}A}
    \\
    &=\frac{1}{2\kappa}\int_\M\eta^{\mu\nu}\wedge\left(\Omega_{\mu\nu}-A\indices{_\mu_\rho}\wedge A\indices{^\rho_\nu}\right)+
    \frac{\varepsilon}{2\kappa}\int_{\partial\M}\left.\left(\tilde{K}_a+K_a\right)\wedge\eta^{\n a}\right|_{\partial\mathcal{M}}
    \mdot
\end{split}\end{align}

At this point, we have finished the first part of our derivation of (S)TEGR. In the second part, we must enforce teleparallelism by setting $\Omega_{\mu\nu}=0$. Naively setting the curvature 2-form to zero would leave us with a bulk term quadratic in $A_{\mu\nu}$ and a boundary term depending on the extrinsic curvature. However, a more careful treatment of the boundary term reveals that it needs to be eliminated as well.

To see this, let us recap a few details regarding GHY terms which have been discussed in~\cite{hess2022}. For the gravity actions~$S$ we consider, taking variations results in $\delta S=\delta S_\mathrm{eom}+\delta S_\mathrm{bdy}$, where $\delta S_\mathrm{bdy}$ is a boundary contribution. Enforcing Hamilton's principle, we expect $\delta S=0$ to yield the equations of motion of a system, but this is spoiled by  $S_\mathrm{bdy}$. We resolve this by including the GHY term~$S_\mathrm{GHY}$ to the action constructed such that $\delta S_\mathrm{GHY}=-\delta S_\mathrm{bdy}$ and $0=\delta (S+S_\mathrm{GHY})=\delta S_\mathrm{eom}$ is enforced. The crucial point in the above discussion is that a non-trivial $\delta S_{\rm bdy}$ stems \textit{solely} from the variation with respect to curvature.  In other words, if we consider a generic action $\mathcal{S}$  constructed out of the torsion and non-metricity fields alone, $\delta \mathcal{S}=\delta \mathcal{S}_\mathrm{eom}$ is already satisfied. The addition of a GHY term would actually spoil the well-definedness of the variational principle, unless its variation is tuned to vanish identically, since it would result in $\delta(\mathcal{S}+S_\mathrm{GHY})=\delta \mathcal{S}_\mathrm{eom}-\delta S_\mathrm{bdy}$.

Thus, in treating the boundary term of~\eqref{eq_eh_action_transformed}, we must first separate it into the GHY term for curvature and remaining boundary part. Then and only then we can consistently take the teleparallel limit by setting both the curvature and its corresponding GHY term to zero. For~\eqref{eq_eh_action_transformed}, the entire boundary term is in fact the GHY term due to curvature~\cite{hess2022}. Therefore, the telepallel limit for the Einstein-Hilbert action consists of setting $\Omega_{\mu\nu}=0$ and eliminating the boundary term. Hence, the \textit{complete and well-defined} action for (S)TEGR is

\begin{align}\label{eq_final_action}
    S^\mathrm{(S)TEGR}
    =-\frac{1}{2\kappa}\int_\M\eta^{\mu\nu}\wedge A\indices{_\mu_\rho}\wedge A\indices{^\rho_\nu}\mdot
\end{align}
The action $S^{\rm (S)TEGR}$ coincides with the differential form (S)TEGR Lagrangian found in~\cite{nester1998} and the components of~\eqref{eq_final_action} also reproduce the (S)TEGR bulk actions considered in e.g.~\cite{lucas2008,aldrovandi2013,jimenez2019}. It is particularly interesting to note that the components of the Lagrangian in~\eqref{eq_final_action} are
\begin{align}\label{eq_local_5}
    \eta^{\mu\nu}\wedge A_{\mu\rho}\wedge A\indices{^\rho_\nu}
    =
    2\sqrt{|g|}\mathrm{dVol}_\mathcal{M}\,A\indices{^\mu_\rho_{[\mu|}}A\indices{^\rho^\nu_{|\nu]}}\mcomma
\end{align}
which also have been found to be the components of 
\begin{align}\label{eq_local_4}
    T_\mu\wedge \star T^\mu
\end{align}
in~\cite{lucas2008,aldrovandi2013} for 3+1d TEGR. However, $\star$ denotes a generalized Hodge dual which is necessary for soldered bundles and a gauge interpretation of the TEGR Lagrangian. The generalization of this Hodge dual to arbitrary dimensions is involved and not known so far~\cite{huang2014}. Hence, it is not immediately clear what the component expression of~\eqref{eq_local_4} is in general dimensions. In contrast to that, our expression~\eqref{eq_local_5} has well-defined and directly accessible components in an arbitrary number of dimensions.  Apart from the bulk dynamics of (S)TEGR, our formalism also establishes unambiguously the (S)TEGR GHY term. We find it to vanish identically. This is the first main result of our paper. 

Our second main result is a new systematic interpretation of the boundary term. Namely, we see that the (S)TEGR boundary term~\eqref{eq_Dc_action} is \textit{not} a GHY term, but rather the difference between the GHY term of GR and the one of metric-affine gravity! That is, \footnote{Note that a more restricted equality between the variations of $S^\mathrm{\overcirc{D}A}$ and $S^{\mathrm{EH}}_{\vphantom{\overcirc{Omega}}\Omega,\mathrm{GHY}}$  has been proven in \cite{PhysRevD.96.044042}.}
\begin{align}
    S^\mathrm{\overcirc{D}A}=S^{\mathrm{EH}}_{\vphantom{\overcirc{Omega}}\Omega,\mathrm{GHY}}-S^\mathrm{EH}_{\overcirc{\Omega},\mathrm{GHY}}\mdot
\end{align}

As a final comment we note that while our derivation is in the language of differential forms, it may equally well be reformulated in a more traditional way in terms of tensor components. We show this explicitly in appendix~\ref{sec_tensor}.

\section{Generalization to \texorpdfstring{$\mathcal{L}(\overcirc{\Omega}\indices{^\mu_\nu})$}{f(curvature)} theories of gravity}
\label{sec_generalization}

In the present section, we will go beyond the Einstein-Hilbert theory of gravity and consider whether general extensions of GR built out of $\overcirc{\Omega}\indices{^\mu_\nu}$ are equivalent to extensions of (S)TEGR. In this way, we derive a generalized geometrical trinity (GGT) of gravity beyond the linear Lagrangian regime. 

Consider a generic action depending only on Riemannian curvature,
\begin{align}\label{eq_action_curv}
    S_{\overcirc{\Omega}}=\int_\mathcal{M}\mathcal{L}(\overcirc{\Omega}\indices{^\mu_\nu}) + S_{\overcirc{\Omega}, {\rm GHY}}\mcomma
\end{align}
in which we do not fix the form of the Lagrangian~$\mathcal{L}$. The GHY term corresponding to ${\cal L}$ was derived in~\cite{hess2022} and takes the form
\begin{align}\label{eq_action_curv_GHY}
    S_{\overcirc{\Omega}, {\rm GHY}}
    =
    \varepsilon \int_{\partial\mathcal{M}}\left. \overcirc{K}^a\wedge\hodge(\varphi_{\n a}- \varphi_{a\n})\right|_{\partial\mathcal{M}}\mcomma
\end{align}
where $\hodge \varphi\indices{_\mu^\nu}$ is defined via $\hodge\varphi\indices{_\mu^\nu}\wedge\delta\overcirc{\Omega}\indices{^\mu_\nu}=\delta_{\overcirc{\Omega}\indices{^\mu_\nu}}\mathcal{L}(\overcirc{\Omega}\indices{^\mu_\nu})$~\footnote{Formally, we have $\hodge\varphi\indices{_\mu^\nu}\wedge\delta\overcirc{\rho}\indices{^\mu_\nu}=\delta_{\overcirc{\rho}\indices{^\mu_\nu}}\mathcal{L}(\overcirc{\rho}\indices{^\mu_\nu})$, see~\cite{hess2022}. For simplifying the presentation we use the equation on-shell, where $\overcirc{\rho}\indices{^\mu_\nu}=\overcirc{\Omega}\indices{^\mu_\nu}$. 
}.
The action built out of the curvature $\Omega\indices{^\mu_\nu}$ reads
\begin{subequations}
    \begin{align}
        &S_{\vphantom{\overcirc{Omega}}\Omega}=\int_\mathcal{M}\mathcal{L}(\Omega\indices{^\mu_\nu}) + S_{\vphantom{\overcirc{Omega}}\Omega, {\rm GHY}}\mcomma
        \\\label{eq_universal_GHY}
       & S_{\vphantom{\overcirc{Omega}}\Omega, {\rm GHY}}=-\int_{\partial\mathcal{M}}\left.\left(  
        -\varepsilon\tilde{K}_a\wedge\hodge\varphi^{\n a}+\varepsilon K^a\wedge\hodge\varphi_{a\n}+\onehalf Q_{\n\n}\wedge\hodge\varphi_{\n\n}
        \right)\right|_{\partial\mathcal{M}}\mdot
    \end{align}
\end{subequations}
To simplify the derivation of the GGT, we use the trick of~\cite{hess2022,deruelle2009} and introduce Lagrange multipliers $\overcirc{\rho}\indices{^\mu_\nu}$ into $S_{\overcirc{\Omega}}$. These allow us to linearize the terms in the action depending on the curvature, namely~\footnote{In~\eqref{eq_local_3}, we have abused notation and denote $S_{\overcirc{\Omega}}$ as well as its version including Lagrange multipliers by the same symbol. The actions are only equal, however, when the equations of motion of $\hodge \varphi\indices{_\mu^\nu}$ are enforced.}
\begin{align}\begin{split}\label{eq_local_3}
    S_{\overcirc{\Omega}}
    &=\int_\mathcal{M}\left(\mathcal{L}(\overcirc{\rho}\indices{^\mu_\nu})+\hodge\varphi\indices{_\mu^\nu}\wedge(\overcirc{\Omega}\indices{^\mu_\nu}-\overcirc{\rho}\indices{^\mu_\nu})\right)
    +S_{\overcirc{\Omega},\mathrm{GHY}}
    \\
    &=\int_\mathcal{M}\left(\mathcal{L}(\overcirc{\rho}\indices{^\mu_\nu})+\hodge\varphi\indices{_\mu^\nu}\wedge(\Omega\indices{^\mu_\nu}-\overcirc{D}A\indices{^\mu_\nu}-A\indices{^\mu_\rho}\wedge A\indices{^\rho_\nu}
    -\overcirc{\rho}\indices{^\mu_\nu})\right)
    +S_{\overcirc{\Omega},\mathrm{GHY}}\mcomma
\end{split}\end{align}
where in the last equality we used the curvature decomposition~\eqref{eq_curvature_split}. Given the linearization of the action, the derivation of the GGT follows as in the previous section with $\eta^{\mu\nu}$ replaced by $\hodge \varphi\indices{_\mu^\nu}$. The equivalent of $S^{\overcirc{D}A}$ in~\eqref{eq_Dc_action_1st_def} in the present case yields
\begin{align}
    \begin{split}
      S^{\overcirc{D}A}_{\varphi}=  -\int_\mathcal{M}\hodge\varphi\indices{_\mu^\nu}\wedge\overcirc{D}A\indices{^\mu_\nu}
        &=
        -\int_{\partial\mathcal{M}}\left. A\indices{^\mu_\nu}\wedge\hodge\varphi\indices{_\mu^\nu}\right|_{\partial\mathcal{M}}-\int_\mathcal{M}A\indices{^\mu_\nu}\wedge\overcirc{D}\hodge\varphi\indices{_\mu^\nu}\mdot
    \end{split}
\end{align}
We expand the $\partial \M$ term in $S^{\overcirc{D}A}_{\varphi}$ in terms of boundary components by using the index decomposition~\eqref{eq_delta_split}, as well as the decomposition~\eqref{eq_c_decomposition} of~$A\indices{^\mu_\nu}$. We arrive at 
\begin{align}
    \begin{split}\label{eq_DA_action_decomposition}
        S^{\overcirc{D}A}_{\varphi}
        &=
        -\varepsilon\int_{\partial\mathcal{M}}\left.\overcirc{K}^a\wedge\hodge(\varphi_{\n a}-\varphi_{a\n})\right|_{\partial\mathcal{M}}
        \\&\hspace{13pt}
        -\int_{\partial\mathcal{M}}\left.\left(  -\varepsilon\tilde{K}_a\wedge\hodge\varphi^{\n a}+\varepsilon K^a\wedge\hodge\varphi_{a\n}+\onehalf Q_{\n\n}\wedge\hodge\varphi_{\n\n} \right)\right|_{\partial\mathcal{M}}
        \\&\hspace{13pt}
        -\int_{\partial\mathcal{M}}\left.A^{ab}\wedge\hodge\varphi_{ab}\right|_{\partial\mathcal{M}}
        -\int_\mathcal{M}A\indices{^\mu_\nu}\wedge\overcirc{D}\hodge\varphi\indices{_\mu^\nu}
        \\
        &=
        -S_{\overcirc{\Omega}, {\rm GHY}}
        +S_{\vphantom{\overcirc{Omega}}\Omega, {\rm GHY}}
        -\int_{\partial\mathcal{M}}\left.A^{ab}\wedge\hodge\varphi_{ab}\right|_{\partial\mathcal{M}}
        -\int_\mathcal{M}A\indices{^\mu_\nu}\wedge\overcirc{D}\hodge\varphi\indices{_\mu^\nu}\mdot
    \end{split}
\end{align}
The result \eqref{eq_DA_action_decomposition} for $S^{\overcirc{D}A}_{\varphi} $ is formally the same as~\eqref{eq_boundary_action_intermediate} after the replacement $\eta_{\mu\nu}\rightarrow \hodge \varphi_{\mu\nu}$ has been made. Thus, whenever the final two terms in \eqref{eq_DA_action_decomposition} vanish identically, we can apply our argument for the geometrical trinity in Einstein gravity mutatis mutandis. Explicitly, in that case we have $S^{\overcirc{D}A}_{\varphi} = -S_{\overcirc{\Omega}, {\rm GHY}} +S_{\vphantom{\overcirc{Omega}}\Omega, {\rm GHY}}$ so that
\begin{align}
\label{eq_GGT'}
    S_{\overcirc{\Omega}}
    =\int_\mathcal{M}\mathcal{L}(\overcirc{\Omega}\indices{^\mu_\nu})
    +\varepsilon \int_{\partial\mathcal{M}}\left. \overcirc{K}^a\wedge\hodge(\varphi_{\n a}- \varphi_{a\n})\right|_{\partial\mathcal{M}}
    =\int_\mathcal{M}\mathcal{L}(-A\indices{^\mu_\rho}\wedge A\indices{^\rho_\nu})\equiv S_A\mcomma
\end{align}
where we enforced teleparallelism by setting $\Omega\indices{^\mu_\nu}=0$ and eliminating the corresponding GHY term~$S_{\Omega,\mathrm{GHY}}$ for consistency. 
The result~\eqref{eq_GGT'} means that the GR and (S)TEGR actions are equivalent after replacing $\overcirc{\Omega}\indices{^\mu_\nu}\mapsto -A\indices{^\mu_\rho}\wedge A\indices{^\rho_\nu} $ and eliminating the GHY boundary term. In tensor components, the corresponding replacement reads $\overcirc{R}\indices{^\mu_\nu_\alpha_\beta}\mapsto A\indices{^\mu_\rho_\beta} A\indices{^\rho_\nu_{\vphantom{\beta}}_\alpha}-A\indices{^\mu_\rho_{\vphantom{\beta}}_\alpha}A\indices{^\rho_\nu_\beta}$, where the components of the deformation one-form are given by~\eqref{eq_A_components}.

In the general case, however, there is no reason to expect that the boundary terms vanish identically, in fact they don't in four-dimensional Chern-Simons modified gravity for instance. Thus, in this case our Lagrange multiplier method is not convenient for obtaining (symmetric) teleparallel equivalents of GR extensions. We can, however, ascertain a particular algorithm for deriving these extensions for a fixed Lagrangian. This proceeds as follows:
\begin{enumerate}
    \item Start with an action constructed solely from the Riemannian curvature~$\overcirc{\Omega}\indices{^\mu_\nu}$. Add an appropriate GHY term to the action. For generic actions~\eqref{eq_action_curv} the GHY term is given by~\eqref{eq_action_curv_GHY}.
    \item Rewrite $\overcirc{\Omega}\indices{^\mu_\nu}$ in the action by means of equation~\eqref{eq_curvature_split}, which amounts to inserting $\overcirc{\Omega}\indices{^\mu_\nu}=\Omega\indices{^\mu_\nu}-\overcirc{D}A\indices{^\mu_\nu}-A\indices{^\mu_\rho}\wedge A\indices{^\rho_\nu}$.
    \item Impose the vanishing curvature condition $\Omega\indices{^\mu_\nu}=0$ of teleparallel gravity, i.e. impose the gauge $\d \omega^\mu_\nu = -\omega^\mu_\rho\wedge\omega^\rho_\nu$.
    \item As indicated by~\eqref{eq_DA_action_decomposition}, the GHY term~$S_{\vphantom{\overcirc{Omega}}\Omega, {\rm GHY}}$ is always present in the boundary contributions found in step~2. Hence, we may always obtain a well-defined variational problem in the teleparallel limit by subtracting~$S_{\vphantom{\overcirc{Omega}}\Omega, {\rm GHY}}$ from the action.
\end{enumerate}

The additional boundary terms present in~\eqref{eq_DA_action_decomposition} (apart from the GHY contributions) are not necessary for establishing the well-posedness of the variational problem, as the teleparallel gravity action is a function of the first derivatives of the metric and/or coframe. Instead, they seem to define a boundary field theory, whose role is to modify the boundary behaviour of the bulk fields a l\'a \cite{SYMANZIK19811}. Thus, we posit that the additional boundary terms define a boundary field theory that is necessary for matching the degrees of freedom between GR and (S)TEGR extensions by enforcing the appropriate boundary conditions on the bulk fields. We leave the analysis of this question for particular GR models for future work.

In conclusion, we find that a generalization of the geometrical trinity of gravity can also be defined between extensions of GR and (S)TEGR, which can be calculated using our prescribed algorithm. This constitutes what we call the generalized geometrical trinity of gravity. Note that in the appropriate limit our results reproduce those of~\cite{percacci2022}. Beyond the theories considered here, one may investigate theories of gravity built on the irreducible components of curvature, torsion or non-metricity with respect to $\mathrm{SO}(1,n-1)$~\cite{hehl1995}. In general, these theories have different symmetries and, hence, different degrees of freedom and there is a~priori no reason to expect any equivalence relation between them. In this sense, our work may be used as a guideline for building (S)TEGR theories that may, even in principle, make predictions differing from those of GR. 

\section{Conclusions and outlook}\label{sec_conclusions}

In this work, we have employed the 3+1 decomposition of the connection within metric-affine gravity to elucidate the precise role of boundary terms appearing in the derivation of the geometrical trinity of gravity. We have further established that the actions of TEGR and STEGR are in no need of GHY terms in order to admit a well-defined variational formulation. In particular, we showed that the boundary term known in the literature for TEGR and STEGR is a difference of GHY terms. Furthermore, we have generalized the geometrical trinity of gravity by showing that any extension of GR built on the curvature two-form $\overcirc{\Omega}\indices{^\mu_\nu}$ can be mapped to an extension of (S)TEGR. We provide an explicit four-step algorithm on how to arrive at said extension at the end of section \ref{sec_generalization}. Said extensions also exhibit non-trivial boundary terms, whose role in this case is not to ameliorate an ill-defined variation. Rather they enforce, in general, non-trivial boundary conditions on the bulk fields.

In view of our results, it will be interesting to re-evaluate existing results, which were derived including the GHY term of GR to the (S)TEGR Lagrangian. This will influence how we understand the thermodynamic properties of black holes within (S)TEGR, which were previously shown to be equivalent to those of GR precisely because of the included GHY term~\cite{paci2021,PhysRevD.96.044042}. For example, we may apply Wald's formalism to TEGR, which relies solely on the assumption of diffeomorphism invariance, by adapting the results of   \cite{Heisenberg:2022nvs} for STEGR. This is particularly important in view of applying the fluid/gravity correspondence to torsionful spacetimes, since the zeroth order contribution in the hydrodynamic derivative expansion is defined entirely in terms of thermal equilibrium~\cite{kovtun2012,Matthaiakakis:2021spp}. As a precursor to establishing fluid/gravity duality for spin currents, it is necessary to establish holographic renormalization \cite{bianchi2001,deHaro:2000vlm,Skenderis:2002wp} for gravity theories with torsion, for which the results presented here are expected to be useful.  In addition, it will be of interest to re-examine theories of gravity which consider generalized Lagrangians depending on the torsion/non-metricity scalar as well as the GHY term in non-linear fashion. In particular, it will also be of interest to check whether the reformulation of modified theories of gravity in~\cite{boehmer2021} in terms of a non-diffeomorphism invariant decomposition of curvature is equivalent to the construction presented here. As a final point in this regard, we can  use the (S)TEGR action~\eqref{eq_final_action} in order to derive a closed form for the generalized Hodge duals of torsion and non-metricity, defined in~\cite{lucas2008,huang2014,aldrovandi2013}, for arbitrary dimension. 

Furthermore, the generalized geometrical trinity of gravity has interesting implications for string theory. Namely, our proof of the equivalence may be applied mutatis mutandi to theories of gravity obtained from string theoretic constructions, after the matter fields have been consistently integrated out of the low-energy spectrum. This suggests we may reformulate the graviton dynamics within string theory in terms of torsion or non-metricity degrees of freedom. It would be interesting to carry out this reformulation explicitly in textbook string and supergravity theories~\cite{polchinski_1998,becker_becker_schwarz_2006,Kiritsis:2019npv,Freedman:2012zz}. Including matter fields in the above reformulations is also an interesting, but challenging problem since enforcing supersymmetry leads to non-trivial constraints on the connection~\cite{Bejarano:2022rqh}.

An additional interesting venue of research are topological theories of gravity. In view of our generalized geometrical trinity of gravity, the existence of these topological theories of gravity expressed in terms of curvature implies the existence of topological theories of torsion or non-metricity. It would be interesting to derive and examine the properties of such topological theories. We note that a first step in this direction has already been taken for Gauss-Bonnet gravity, within the setting of teleparallel gravity. Namely, the authors of~\cite{Kofinas:2014owa} have used the Gauss-Bonnet Lagrangian to derive a topological invariant based entirely on torsion. In addition, recently the symmetric teleparallel extension of Gauss-Bonnet gravity has been worked out in \cite{Bajardi:2023gkd}.  Similarly, such invariants should exist as well for topological actions containing higher powers of the curvature two-form.   

Finally, note we considered the cases of spacelike and timelike boundaries. However, spacetimes may also have boundaries with a lightlike normal vector, $n_\mu n^\mu = 0$. These are particularly important from the point of view of gauge/gravity duality, since hypersurfaces with lightlike  boundaries are used to compute quantum complexity~\cite{PhysRevLett.116.191301, Couch2017, Belin:2022xmt}. Furthermore, lightlike hypersurfaces are important in spacetimes containing black holes, since the black hole horizons may be interpreted as a lightlike boundary. Hence, it would be interesting to extend our analysis to the case of lightlike boundaries based on the formalisms in~\cite{Gourgoulhon:2005ng,Lehner:2016vdi}. 

\medskip
\begin{center}
\large{\textbf{{Acknowledgements}}}
\end{center}
\medskip

We thank Lavinia Heisenberg, Christian Peifer, Martin Krššák, Tomi Koivisto and Erik Jensko for discussions and useful comments on this paper. We are grateful for the funding by the Deutsche Forschungsgemeinschaft (DFG, German Research Foundation) through Project-ID 258499086—SFB 1170 \enquote{ToCoTronics}, through the Würzburg-Dresden Cluster of Excellence on Complexity and Topology in Quantum Matter – ct.qmat Project-ID 390858490—EXC 2147, and in part through the German-Israeli Project Cooperation (DIP) grant \enquote{Holography and the Swampland}. The research of B.H. is funded by the Bundesministerium für Bildung und Forschung (BMBF, German Federal Ministry of Education and Research) through the Cusanuswerk - Bischöfliche Studienförderung. I.M.'s research is supported by the \enquote{Curiosity Driven Grant 2020} of the University of Genova and by the INFN Scientific Initiatives SFT: \enquote{Statistical Field Theory, Low-Dimensional Systems, Integrable Models and Applications} and by the STFC consolidated grant (ST/X000583/1) \enquote{New Frontiers In Particle Physics, Cosmology And Gravity}.

\appendix

\section{Tensor component formulation of the (S)TEGR results}\label{sec_tensor}
The results of section~\ref{sec_main} translate to the familiar tensor component formulations of (S)TEGR qualitatively. The transformation of the Einstein-Hilbert action to (S)TEGR may be found in many references, see~\cite{bahamonde2021,jarv2018} for example. We will therefore not repeat the derivation at this point. However, the connection of the (S)TEGR boundary term to the GHY term proceeds different as in differential form language. Since the qualitative results coincide with those presented in section~\ref{sec_main}, we only give their derivation in tensorial language for TEGR. They generalize to STEGR similarly.

We adapt the notations from~\cite{bahamonde2021}. Recall that the Riemannian Ricci tensor $\overcirc{R}$ transforms to the torsion scalar~$T$ and the TEGR boundary term $B\defeq2\overcirc{\nabla}_\mu T\indices{^\nu_\nu^\mu}$ as
\begin{align}
    \overcirc{R}=R-T+B\mdot
\end{align}
To convert the integral of $B$ into a boundary term, we invoke Stokes' theorem and obtain
\begin{align}\label{eq_local_2}
    \int_{\mathcal{M}} \d\mathrm{Vol}_{\mathcal{M}}\sqrt{|g|}B=\int_{\partial \mathcal{M}} \d\mathrm{Vol}_{\partial \mathcal{M}} \sqrt{|\gamma|}2\varepsilon n^\mu T\indices{^\nu_\nu_\mu}\mdot
\end{align}
Showing the connection of the TEGR boundary term and the GHY term thus is the same as performing the 3+1 decomposition of~$n^\mu T\indices{^\nu_\nu_\mu}$ which we do next. Decomposing the indices of the torsion tensor by means of~\eqref{eq_delta_split} yields
\begin{align}
    n^\mu T\indices{^\nu_\nu_\mu}=e^a_\nu e_a^\rho n^\mu T\indices{^\nu_\rho_\mu}
\end{align}
since $n_\mu n^\nu n^\rho T\indices{^\mu_\nu_\rho}$ vanishes due to the antisymmetry of $T\indices{^\mu_\nu_\rho}$. From~\cite{hess2022} we use the 3+1 decomposition of torsion in the form
\begin{align}
    e^a_\mu T^\mu=T^a+N K^a\wedge\phi\mcomma
\end{align}
where $T^a\equiv D\phi^a$ is the boundary torsion two-form and $\phi$ and $N$ have been defined in~\eqref{eq_g_decomposition}. This decomposition implies that its tensor components transform to the boundary as
\begin{align}
    e^a_\nu e_b^\rho n^\mu T\indices{^\nu_\rho_\mu}=e^\rho_b n^\mu T\indices{^a_\rho_\mu}+ K\indices{^a_b}\mdot
\end{align}
The Riemannian limit of the latter equation yields
\begin{align}
    e^\rho_b n^\mu T\indices{^a_\rho_\mu}=- \overcirc{K}\indices{^a_b}\mdot
\end{align}
We insert these results into~\eqref{eq_local_2} to obtain
\begin{align}
    \int_\mathcal{M} \d\mathrm{Vol}_{\mathcal{M}}\sqrt{|g|}B
    =
    2\varepsilon  \int_{\partial\mathcal{M}} \d\mathrm{Vol}_{\partial \mathcal{M}} \sqrt{|\gamma|}\left(K\indices{^a_a}-\overcirc{K}\indices{^a_a}\right) \mdot
\end{align}
This is the difference between the GHY term of the Lagrangian~$\overcirc{R}$ and the one of $R$ in components. Analogous to the differential geometric formulation in section~\ref{sec_main} we transition from the GR action to the TEGR action in components by setting $R=0$. Since the GHY term is there to cancel the boundary terms which are non-vanishing under variation of the action, it introduces non-vanishing boundary terms in the variational calculation if $R=0$ is imposed. In this way it makes the variational problem ill-defined again. Thus, we need to cancel the GHY term in the transition to TEGR additional to imposing~$R=0$. Hence, we find the same behavior as in the differential form calculation in section~\ref{sec_main}. Of course, this result also implies that a GHY term must be included in the Riemannian Einstein-Hilbert action before transforming to (S)TEGR. Hence, the transformed Einstein-Hilbert action in tensor components reads
\begin{align}\begin{split}
    S^\mathrm{EH}&=\frac{1}{2\kappa}\int_\mathcal{M}\mathrm{dVol}_\mathcal{M}\sqrt{|g|} \overcirc{R} + \frac{\varepsilon}{\kappa}\int_{\partial\mathcal{M}}\mathrm{dVol}_{\partial\mathcal{M}}\sqrt{|\gamma|}\overcirc{K}\indices{^a_a}
    \\&=
    \frac{1}{2\kappa}\int_\mathcal{M}\mathrm{dVol}_\mathcal{M}\sqrt{|g|} (R -T+B)+ \frac{\varepsilon}{\kappa}\int_{\partial\mathcal{M}}\mathrm{dVol}_{\partial\mathcal{M}}\sqrt{|\gamma|}\overcirc{K}\indices{^a_a}
    \\
    &=
    \frac{1}{2\kappa}\int_\mathcal{M}\mathrm{dVol}_\mathcal{M}\sqrt{|g|} (R -T)+ \frac{\varepsilon}{\kappa}\int_{\partial\mathcal{M}}\mathrm{dVol}_{\partial\mathcal{M}}\sqrt{|\gamma|}K\indices{^a_a}\mdot
\end{split}\end{align}
TEGR is therefore completely described by the action
\begin{align}\label{eq_TEGR_action_components}
    S^\mathrm{TEGR}=-\frac{1}{2\kappa}\int_\mathcal{M}\mathrm{dVol}_\mathcal{M}\sqrt{|g|} \,T\mdot 
\end{align}
This result proves that~\eqref{eq_TEGR_action_components} is the full TEGR action. Schematically, \eqref{eq_TEGR_action_components} means that the GHY term needs to be canceled in the transition from general relativity to TEGR while no TEGR boundary term~$B$ is needed. The results of the tensor component formulation of TEGR thus coincide with the differential form formulation in section~\ref{sec_main}.

\bibliography{bibliography.bib}

\end{document}